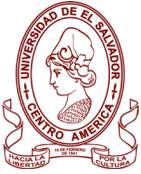

# Revista MINERVA

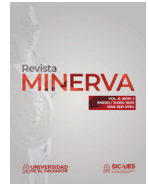

Artículo Científico | Scientific Article

# Aplicación de tecnologías IoT en el control y seguimiento de trasporte de carga terrestre

## Application of IoT technologies in the control and monitoring of road freight transport

Omar Otoniel Flores-Cortez[1]

Bruno Gonzales Crespin[2]

Correspondencia
omar.flores@ues.edu.sv

Presentado: 9 de diciembre de 2022
Aceptado: 20 de febrero de 2023

1. Maestría en Ingeniería para la Industria con Especialización en Telecomunicaciones, Escuela de Posgrado Facultad de Ingeniería y Arquitectura, Universidad de El Salvador, ORCID: 0000-0003-1754-4090
2. Smart Metrics El Salvador. ORCID: 0000-0002-4066-1298

**RESUMEN**

El transporte de carga vía terrestre es una parte importante en la cadena de suministro del comercio en Latinoamérica. El control y seguimiento de esta actividad es vital para un flujo eficiente y sin pérdidas económicas. La mayoría de los problemas son pérdidas por cambios en el peso de la carga útil a transportar o pérdidas de combustible/tiempo por cambios, caprichosos del conductor, en la ruta programada. Este trabajo tiene como objetivo demostrar el uso de técnicas de Internet de las cosas (IoT) para proponer un prototipo de telemetría para monitorear en tiempo real el peso y la ubicación de un camión de carga, y convertirse en una herramienta tecnológica que soporte tareas de logística de control que conlleven a minimizar las pérdidas económicas. El desarrollo de este proyecto se basó en el modelo de referencia de la arquitectura IoT. En el diseño electrónico la estación IoT se usó un microcontrolador Atmega32u4 junto con un módulo SIM808 GSM y GPS como componente principal. Además, como plataforma de almacenamiento y presentación IoT se utilizaron herramientas de *Amazon Web Services* (AWS). El principal resultado fue un prototipo de un sistema de telemetría para rastrear un vehículo de mediano tonelaje, los datos de peso y posición son accesibles desde cualquier dispositivo con acceso a internet. Las pruebas de campo preliminares han sido satisfactorias y han demostrado que el sistema propuesto es una opción eficiente y de bajo costo para el monitoreo de la posición global y del nivel de carga del camión. La primera etapa del proyecto, se enfocó en el diseño y construcción de un prototipo de IoT para la obtención remota de datos del camión en tiempo real, queda para próximas etapas del proyecto escalar a más nodos y expandir los tiempos de recolección de datos, que servirán para realizar estudios para verificar posibles efectos lógicos y económicos, además de realizar pronósticos con los datos recogidos y así realizar propuestas sobre mejoras en la logística, verificación de rutas, kilometraje y gasto de combustible.

**Palabras clave:** Internet de las cosas, sensores, telemetría, microcontrolador, GPS, transporte de carga

## ABSTRACT

Freight transport of goods and raw materials is a main part of the supply chain in the commercial exchange in Latin America. Control and monitoring of this activity are vital for an efficient economic flow and more importantly without losing money. Most of the problems that generate economic losses occur in cargo freight by land. Losses due to changes in the weight of the payload to be transported or fuel/time losses due to capricious changes by the driver on the scheduled route. This work aims to demonstrate use of Internet of Thing (IoT) techniques to propose a prototype of a telemetry system to monitor in real-time the payload weight and location of a cargo truck and become a technological tool that supports the tasks of monitoring and control of the use of cargo trucks, and together with other logistics measures, leads to minimizing economic losses. The development of this project was based on the IoT architecture reference model: an ATmega32u4 microcontroller was used together with a SIM808 GSM and GPS module as the main component of the IoT Node. In addition, Amazon Web Services (AWS) tools were used as an IoT web platform and cloud data storage. The main result was a prototype of a telemetry system to track a cargo truck via the web, the weight and position data are accessible from any device with internet access through a website. Preliminary field tests have been successful and have shown the proposed system to be an efficient and low-cost option.

## INTRODUCCIÓN

En Latinoamérica, el intercambio de mercancías y materias primas se realiza, principalmente, a través del transporte terrestre. Su ventaja es el uso de infraestructura vial universal y su costo relativamente bajo. Es muy utilizado para distancias inferiores a 1,000 km y cargas inferiores a 44 toneladas. Así, cada región o país cuenta con una gran flota de camiones de carga y una legislación propia para esta importante actividad económica. Sin embargo, a pesar de su amplio uso, el transporte de mercancías es afectado por problemas que se traducen en pérdidas económicas. (Moral, 2014) (Quijano, 2018). Un problema es la falta de control del peso de la carga útil en los camiones, desde que se carga en el muelle de origen hasta que se entrega al cliente. Esto crea inconsistencias entre lo que se envía y lo que se entrega, lo que resulta en una pérdida de dinero para el transportista. Se estima que estas discrepancias de peso se deben a factores como el robo en la carretera, el robo por parte de los conductores, o la falta de normas técnicas en las básculas de los puntos de pesaje a lo largo de la ruta. La sobrecarga en el transporte de mercancías es un problema común que ocurre en todo el mundo. De los efectos negativos que genera la sobrecarga de camiones, quizás el de mayor impacto es el deterioro de las vías, reduciendo su vida útil y generando mayores costos de mantenimiento. Otros efectos negativos son el aumento de la siniestralidad vial, el aumento de las emisiones, las multas y los mayores tiempos de transporte (Carbajal, 2021) (Espinoza, 2018). Otro problema es el gasto desproporcionado de tiempo y combustible durante los viajes, debido a la negligencia de los conductores en cambiar o modificar la ruta establecida, lo que genera retrasos en los tiempos de entrega y producen pérdidas monetarias (Sánchez & Cipoletta Tomassian, 2003) (Beetrack, 2019). Por esta razón, el control y seguimiento de los camiones, a lo largo de la ruta de distribución, se ha convertido en una necesidad, específicamente el seguimiento en tiempo real de variables como el peso de la carga y la posición geográfica del camión, con el fin de tomar acciones para evitar problemas por cambios en ruta o en el peso del camión (Gohin Tay & Vera Bernuí, 2015) (Minero, 2015). Trabajos anteriores han estado relacionados con el desarrollo de sistemas de seguimiento posicional para vehículos (Molina, y otros, 2020). Pero estas se han centrado en el uso de herramientas tecnológicas de alto costo o no están conectadas a un sitio web en tiempo real (Jurado Murillo, y otros, 2020) (Alrifaie, Harum, Othman, Roslan, & Shyaa, 2018). El uso de tecnologías GPS junto con placas de desarrollo para microcontroladores como Arduino es muy común en trabajos anteriores (Cadena, Matamoros, Pérez, & Escobar, 2019) (San Hlaing, Naing, & San Naing, 2019). Otros trabajos se han enfocado en desarrollar dispositivos para monitorear el peso de camiones de carga, sin embargo, también se enfocan en tecnologías con alto presupuesto o que no reportan en tiempo real. Algunos de estos trabajos más antiguos solo se enfocan en monitorear una variable, ya sea el peso o la ubicación de la carga del camión (Putra, y otros, 2019) (Hernandez & Hyun, 2020). En la mayoría de los desarrollos se han utilizado sensores de peso analógicos, estos se basan en la deflexión de los soportes de los amortiguadores de la suspensión del camión, convirtiendo este ángulo de deflexión en un voltaje analógico (Seo, Shin, Lee, Ko, & Tumenjargal, 2021) (Stawska, Chmielewski, Bacharz, Bacharz, & Nowak, 2021), aunque los sensores de peso digitales también se han usado menos en sistemas similares (Radhakrishnan, y otros, 2021) (Oskoui, Taylor, & Ansari, 2020). La arquitectura de un sistema de Internet de las Cosas (IoT) está definida por dos bloques principales: el nodo sensor y la plataforma de Internet. Los nodos sensores son el dispositivo de telemetría equipado con elementos sensores que toman lecturas de diferentes variables de comportamiento del equipo a monitorear (Rosa & Flores-Cortez, 2017). La plataforma de Internet, también llamada nube, es donde se almacenarán los datos recolectados por el Nodo, además de su visualización a través de tableros web (Bahga & Madisetti, 2014) (Lv, y otros, 2017). El enlace entre estos Nodos y la plataforma puede implementarse a través de tecnologías de radio como Wifi, Bluetooth, GSM/GPRS, LoRa, entre otras (Chanchí G, Ospina A, Campo M, & others, 2021) (Golondrino, Alarcón, & Muñoz, 2020) (Saleem, Zeebaree, Zeebaree, & Abdulazeez, 2020). Los microcontroladores Atmega y ESP son la opción más utilizada en la implementación del procesador del nodo sensor (Calixto-Rodriguez, y otros, 2021) (Bento, 2018) (Singh & Kapoor, 2017). Para la plataforma de Internet, las opciones más utilizadas son Amazon Web Services, Google IoT Core, Tingspeak y Ubidots (Hejazi, Rajab, Cinkler, & Lengyel, 2018) (Balakrishna & Thirumaran, 2019).

Este trabajo propone un sistema telemático, en tiempo real, de bajo costo basado en tecnologías IoT. Las estaciones IoT están equipadas con sensores que pueden leer el peso de un camión de carga y las variables de ubicación y enviar los datos a través de un enlace celular GSM/GPRS a Internet y de una plataforma de almacenamiento y despliegue web IoT accesible desde cualquier dispositivo conectado a Internet para que el personal pueda monitorizar y controlar posibles situaciones anómalas en la ruta de transporte.

## METODOLOGÍA

Este trabajo tiene como objetivo demostrar el uso de técnicas de IoT para proponer un prototipo de sistema de telemetría para monitorear en tiempo real el peso de la carga útil y la ubicación de un camión de carga. El desarrollo metodológico de este sistema propuesto se basó en el Modelo de Referencia Arquitectónica de IoT (Bahga & Madisetti, 2014).

### Propósito y especificación de sistema propuesto

Propósito: monitoreo automatizado, del peso y posición de un camión de carga, con enlace celular GSM para reportar en tiempo real a través de un tablero web. Comportamiento: una estación electrónica con sensores capaces de tomar medidas del peso y posición del camión, un controlador digital central programado para realizar lecturas periódicas de sensores y enviar los datos recolectados vía enlace celular GSM/GPRS a la plataforma de Internet. Gestión: el sistema se puede monitorizar a través de Internet y la gestión de la programación y configuración del Nodo Sensor se puede realizar localmente a través de un puerto USB previsto en la propia estación. Análisis de datos: los datos recopilados por el sensor se procesan en la propia estación y luego se envían en valores de carga útil a la nube. Implementación de aplicaciones: el software o firmware de control de la estación permanece dentro de la memoria flash del microcontrolador y se codifica en lenguaje de programación C. Se utiliza una plataforma IoT con paneles de visualización web para monitorear los datos producidos por el nodo. Seguridad: el sistema debe tener autenticación de usuario y un protocolo JSON con autenticación de token para recibir cargas de datos desde la estación a la plataforma. El acceso al panel de datos web será de acceso público *vía* Internet.

### Especificación del proceso del sistema propuesto

Un solo caso de operación en un ciclo repetitivo se define a través del firmware en el controlador digital: cuando el sistema arranca, ejecuta acciones para configurar el hardware interno y externo del microcontrolador, luego lee los sensores de peso y posición, los formatea en protocolo JSON y finalmente los envía a la plataforma IoT a través de una red celular GSM/GPRS, todo este proceso es periódico, especificado en la Figura 1.

### Especificación del modelo de dominio del sistema propuesto

Entidad física: es monitorizar el camión de carga; su peso y posición global actual. Entidad virtual: representa una entidad física en el mundo digital, por lo que solo se define una para el camión de carga. Dispositivo central: controlador digital programable con sensores GPS de posición y peso, con transceptor de red celular GSM. Recurso: firmware que se ejecuta en el dispositivo y un script de configuración que

**Figura 1**

*Especificación del algoritmo principal de la estación o Nodo Sensor dentro del camión*

```
Algorithm 1: Especificacion del Proceso para el Nodo Sensor IoT
   Result: Periodicamente t leer el sensor GPS y sensor de peso dentro
           del camion de carga y enviar via enlace celular las lecturas a la
           Plataforma IoT respectiva.
 1 Configurar hardware interno del Microcontrolador;
 2 Configurar el hardware del trasceptor GSM;
 3 Configurar el hardware del trasceptor GPS;
 4 Definir t ;                              // minutos entre lectura/envio
 5 while True do
 6    leer el sensor GPS ;                  // long, lat, timestamp
 7    leer la salida del sensor analogo de peso ;              // w
 8    formatear el paquete JSON con datos lon, lat, w, time;
 9    conectarse a la red GSM;
10    activar el uso de datos GPRS;
11    realizar un requerimiento HTTP POST hacia la plataforma de
          almacenamiento web;
12    esperar por la respuesta del servidor;
13    if respuesta == 200 then
14        apagar uso de datos GPRS;
15        esperar t;
16    else
17        reintentar t ;                    // reintentar HTTP POST
18    end
19 end
```

se ejecuta en la nube de IoT. Servicio: el servicio de la estación se ejecuta de forma nativa en el dispositivo.

### Especificación de vista funcional del sistema propuesto

Se define grupos funcionales (FG) para los diferentes bloques del sistema IoT. Cada grupo interactúa con instancias definidas en el modelo de dominio o con información relacionada con ellas. FG estación: incluye el microcontrolador, el transceptor GSM, sensores de peso y posición GPS. FG comunicación: los protocolos utilizados son enlace 802.11 vía GPRS, capa de aplicación HTTP y protocolo JSON para enviar la carga útil de datos a la plataforma IoT. FG servicios: solo hay un servicio ejecutándose dentro del servicio de control de la estación IoT. Gestión FG: realizada por el recurso de firmware dentro del microcontrolador. FG seguridad: el mecanismo de seguridad es una credencial de usuario de autenticación para la configuración de la nube de IoT. FG aplicación: interfaz web para monitorear los valores producidos por el nodo IoT está en la "nube" como una página de Internet, ver Figura 2.

**Especificación de vista operativa**

Se definen opciones para el despliegue y operación del sistema IoT. Estación de nodo IoT: los componentes principales son un microcontrolador, un transceptor de red GSM para acceso a Internet, un sensor de peso, un sensor de posición GPS. La interfaz de programación de aplicación (API) para comunicación entre la estación electrónica y la plataforma fue AWS (Amazon Web Services) y los protocolos de comunicación implementados incluyen 802.11, IPV4/6, TCP y HTTP. La interface de visualización web, así como la base de datos de alojamiento se implementaron con los servicios y herramientas de AWS. El sistema incluye un servicio de controlador alojado en el nodo o estación electrónica, escrito en lenguaje de programación C y que se ejecuta como un servicio nativo. La administración del nodo se realiza vía el entorno de desarrollo (IDE) de Arduino y la plataforma se administra por medio del portal de AWS.

**Integración de componentes del sistema propuesto**

Componentes para el Nodo IoT: se utiliza un microcontrolador ATmega32u4 como CPU, el chip SIM808 se utiliza como transceptor para la red celular GSM/GPRS, que también incluye un sensor receptor GPS en el mismo paquete. Como sensor de peso se utiliza el sensor GNOM DP con salida analógica, que se coloca en el eje de suspensión del camión, ver Figura 3. También en un segundo prototipo se utiliza el sensor de peso con salida analógica GNOM DDE, que detecta cambios en la presión de las mangueras de amortiguación del camión.

**Desarrollo de aplicaciones**

Desde el punto de vista de las aplicaciones de software desarrolladas para ejecutar el sistema IoT 1) Firmware del nodo IoT: escrito en lenguaje de programación ANSI C, el programa sigue una estructura de bucle único y tareas específicas que se repiten cíclicamente en un período

**Figura 2**

*Vista general de la arquitectura de bloques funcionales del sistema IoT propuesto*

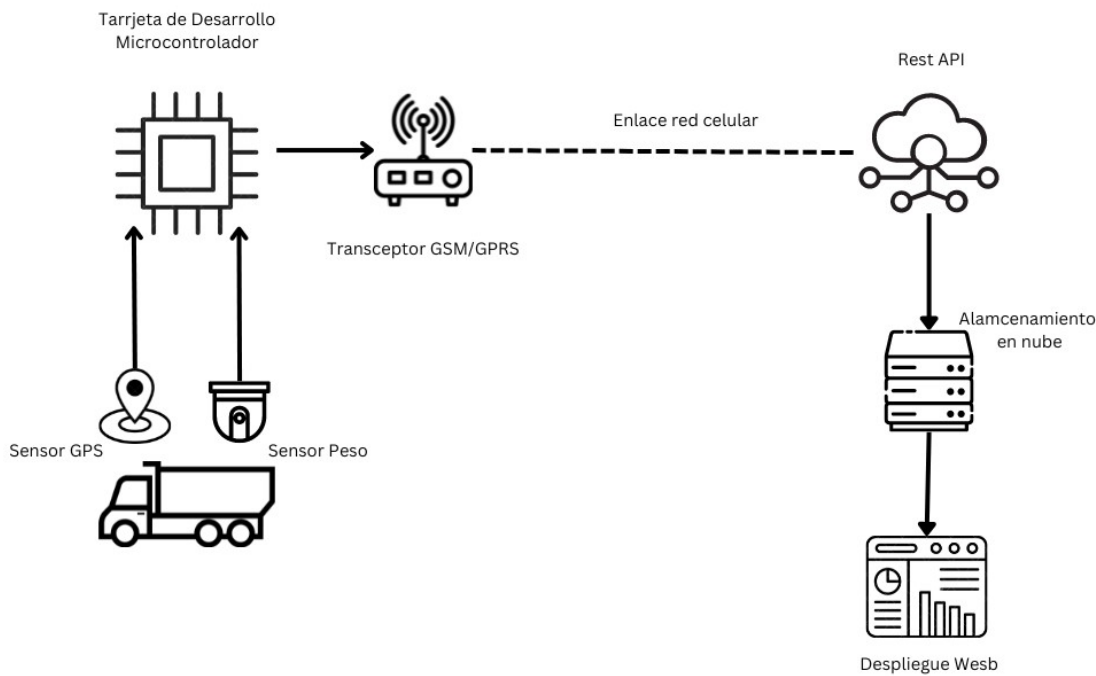

**Figura 3**

*Integración de componentes utilizados en el prototipo del sistema propuesto*

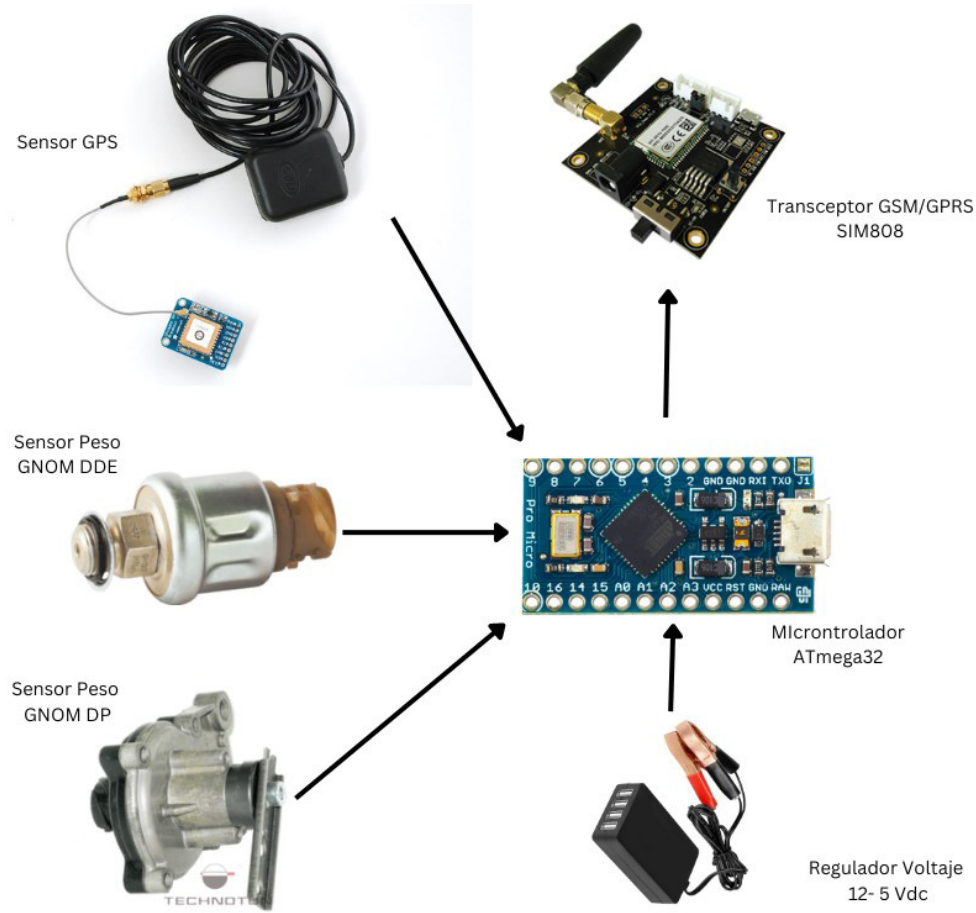

configurable, según Figura q. 2) Configuración del software de la plataforma IoT: script de servicio "Cloud": desarrollado en lenguaje JavaScript alojado en la nube de *Amazon Web Services* (AWS). El protocolo de telemetría JavaScript para la notación de objetos (JSON) se utiliza para enviar y recibir datos entre el nodo de IoT del sensor y la plataforma de IoT. Los servicios de AWS fueron seleccionados por su bajo costo, alta confiabilidad y disponibilidad. Además de tener una curva de aprendizaje relativamente corta. 3) Tablero web: desarrollado con los servicios de alojamiento de AWS, utilizando una caja de herramientas web para configurar el sitio web con tablas de datos y un tablero gráfico para mostrar los datos generados por los sensores.

**RESULTADOS Y DISCUSIÓN**

El principal resultado de este trabajo fue un prototipo de un Sistema IoT para monitorear el peso y la ubicación de un camión de carga en tiempo real.

**Estación de monitoreo de IoT**

Nodo Sensor IoT: estación con sensores electrónicos que permiten tomar medidas de peso y posición GPS de un camión de carga y enviarlas a la plataforma IoT en la nube, ver Figura 4. Como plataforma hardware de desarrollo se utiliza la placa SIM808 GSM/GPRS/GPS IoT del fabricante DFRobot, que incluye el microcontrolador ATmega junto con el SIM808 en la misma placa (DFRobot).

Es un diseño que toma en cuenta las necesidades de las condiciones de la región latinoamericana, basado en componentes electrónicos de última generación, asequibles, eficientes y disponibles en el mercado local. El diseño de la estación de hardware permite agregar más sensores a la estación para aumentar las variables a medir. La estación informa al sitio web dos valores de magnitud detectados cada 10 minutos o se pueden configurar en el firmware del microcontrolador.

La instalación física de la estación es sencilla, se puede empotrar en la estructura de un camión. Los requisitos técnicos para el sitio de instalación son: fuente de alimentación de 12 VDC cerca de la batería del camión y cobertura de Red Celular, la estación está configurada para acceso a Internet a través de un enlace GPRS y utiliza una red celular 2G, ver Figura 5. La puesta en marcha solo requiere definir vía firmware, las credenciales de acceso a la red y una tarjeta SIM con plan de datos activo. En las pruebas de campo se configuró un tiempo de espera entre envíos a la plataforma IoT de 5 a 10 minutos, esto es modificable desde el firmware de la estación.

Entre las características eléctricas del prototipo de estación tenemos: Voltaje de operación: 12 VDC @ 0.4 W máx. Temperatura de funcionamiento: +60 °C máx. Operación de

**Figura 4**

*Ensamble de la estación electrónica para el Nodo Sensor del sistema IoT propuesto*

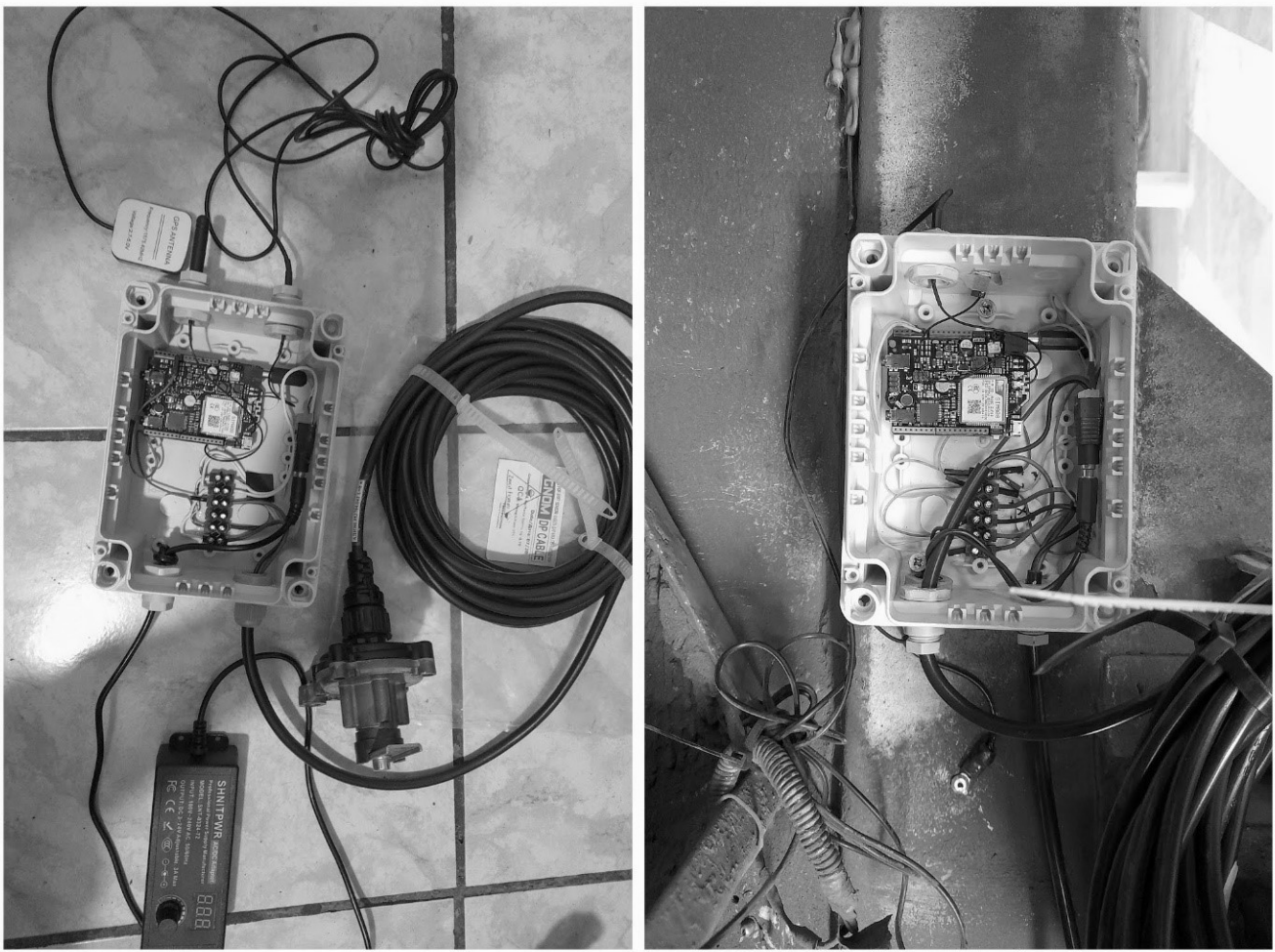

**Figura 5**

*Diferentes pruebas de instalación del nodo sensor dentro de la estructura del camión.*

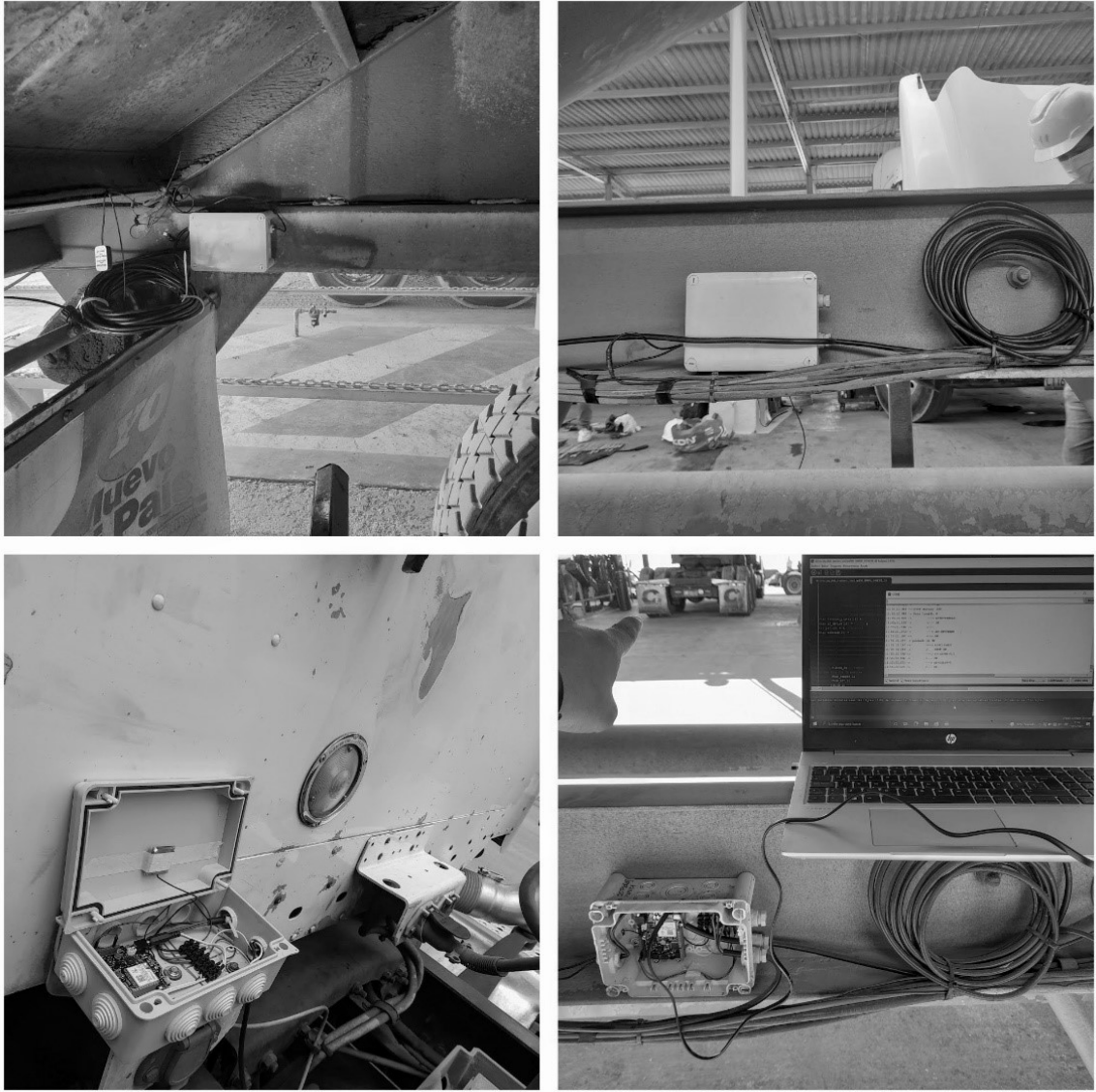

medición: Rango de peso de 1 a 10 toneladas. Localización GPS Precisión de posición horizontal: < 2,5m CEP. Rendimiento de la comunicación: Enlace: Red 2G de banda cuádruple GSM/GPRS. Conectividad GPRS: 85,6 kbps máximo y tarjeta SIM estándar.

**Plataforma web y prueba de campo**

Se realizo una primera etapa de prueba con un Nodo IoT colocados en diferentes camiones de la flota de la empresa CORPORIN S.A de C.V., la cual se enfoca en los servicios de transporte de carga en El Salvador y en algunas ciudades de Guatemala y Honduras (de Corporin SA, 2021). Se diseñó un sitio web accesible desde Internet para el personal de control y logística de la empresa. Este sitio web incluye tablas y tableros para ver el historial de valores de peso, longitud y latitud informados por la estación instalada dentro del camión de carga, ver Figura 6. Los datos presentados corresponden a una prueba durante un recorrido de un camión, de trasporte

**Figura 6**

*Captura del sitio web donde se muestran los datos recolectados por el nodo sensor dentro del camión*

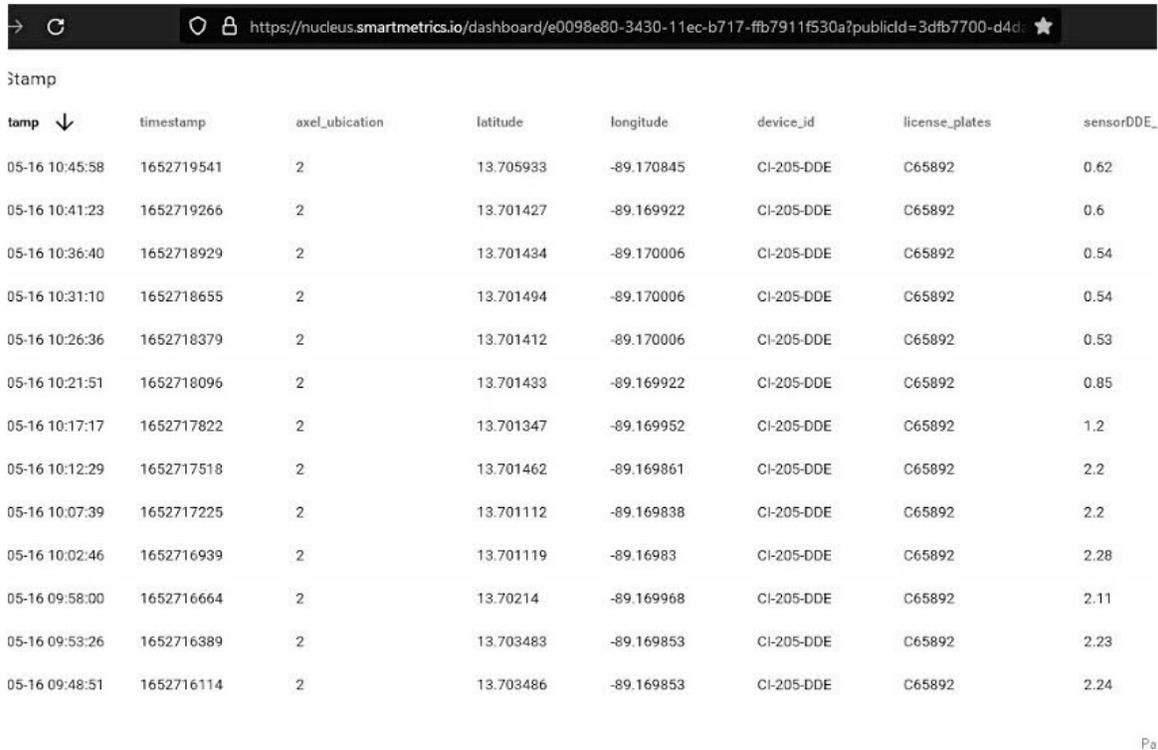

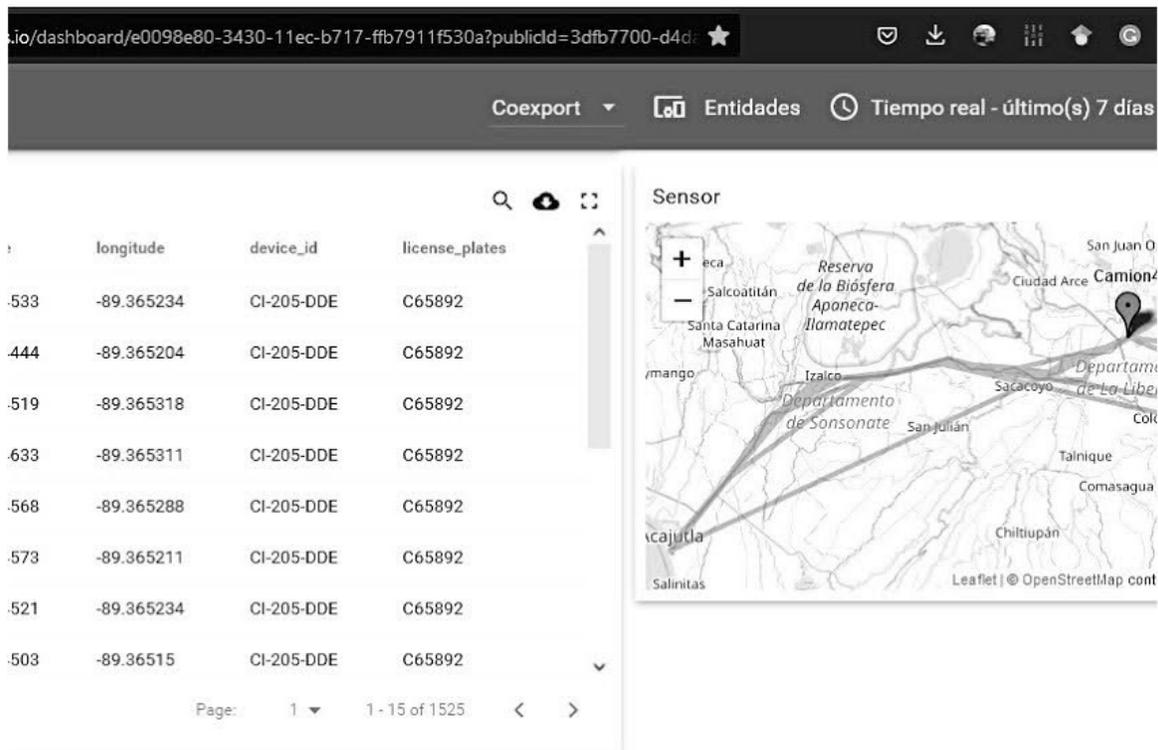

de cemento a granel, desde la planta productora hasta un sitio de almacenaje y embalaje, datos cedidos por CORPORIN S.A de C.V para efectos de esta divulgación.

En esta prueba de campo inicial se instaló un Nodo IoT por un periodo que contemplaba la ruta del camión monitoreado, normalmente de 1 a 2 días. Se seleccionaron camiones con una ruta relativamente corta, Puerto de Acajutla – Opico o Cementera Holcim- Ilopango, principalmente por efecto de poder dar soporte de forma relativamente rápida a algún nodo durante la ruta. El rendimiento del sistema hasta ahora ha sido satisfactorio con respecto al objetivo de poder visualizar remotamente la posición y el nivel de peso del camión bajo análisis. El enlace de telemetría no ha sufrido pérdidas y se ha mantenido estable, se detectaron momentos de perdida de enlace en lugares donde el camión transitaba en zonas de nula recepción de la señal GPS, por ejemplo, bajo puentes o túneles. Se realizaron varias pruebas con diferentes Nodos IoT ubicados en diferentes puntos dentro de la estructura del camión, uno de los mejores resultados de enlace se obtuvo con la estación ubicada detrás de la cabina del camión y con vistas a cielo abierto, además, de que se observó mejoras en el enlace GPS y GSM al utilizar antenas externas colocadas en el techo del camión.

## Discusión

El desarrollo de un sistema IoT, para monitorear en tiempo real el peso y posicionamiento de un camión a lo largo de su ruta, es un paso fundamental en el estudio de comportamiento, impactos y acciones sobre posibles anomalías de datos en la ruta y peso del camión. En este trabajo demuestra el uso efectivo y a bajo costo de técnicas de Internet de las cosas (IoT) para diseñar y construir de sistema telemétrico para monitorear en tiempo real el peso y la ubicación de un camión de carga. El sistema propuesto sirve como una herramienta tecnológica que apoya las tareas de monitoreo y control para el uso de camiones de carga, que, junto con otras medidas logísticas, podrá contribuir a minimizar las pérdidas económicas por inconsistencias en el peso de la carga.

El sistema propuesto fue desarrollado utilizando técnicas de última generación en electrónica, programación e Internet de las cosas, lo que permitió producir equipos de bajo costo que funcionan de acuerdo con los requerimientos esperados. Herramientas como el microcontrolador ATmega junto con el lenguaje de programación ANSI C permiten desarrollar prototipos de IoT eficientes a bajo costo, con tiempos de desarrollo cortos y alto rendimiento. Además, el uso de la caja de herramientas de AWS ha permitido un monitoreo rápido y fácil de la plataforma y el sitio de desarrollo web a los datos desde cualquier dispositivo y en tiempo real. El aporte de este trabajo fue mostrar innovadoras técnicas para el uso de componentes de hardware y software en la implementación de Sistemas IoT. Además de ser una aplicación ad-hoc para la necesidad y contexto del transporte de carga en El Salvador, donde aspectos como el bajo costo y la personalización son valiosos para propuestas tecnológicas innovadoras. Estos se pueden aplicar en nuevos desarrollos, lo que permite la creación rápida y eficiente de prototipos.

Como trabajo a futuro, este proyecto de investigación tiene la tarea de desarrollar más estaciones agregando diferentes sensores para capturar más variables sobre el rendimiento del camión y establecer más validación de prueba de campo. Implementar una plataforma, en la nube, más robusta, con paneles y tablas potentes con datos listos para leer. Adicionalmente, buscamos implementar una red de monitoreo a través de enlaces de radiofrecuencia y aplicar análisis Big Data y generar pronósticos con base a los datos que producen las estaciones. El resultado de este trabajo puede ser utilizado en el desarrollo de nuevas líneas de investigación aplicada, en áreas como el análisis de suelos o acuíferos, la monitorización en campos agrícolas y ganaderos, el análisis del rendimiento deportivo, etc.

En conclusión, en esta primera etapa del proyecto, se ha verificado que es posible implementar sistemas de Internet de las cosas con componentes accesibles y utilizando herramientas Open Hardware de bajo costo. Con las pruebas iniciales realizadas en esta etapa se han obtenido resultados que permiten continuar en el desarrollo del proyecto de monitoreo de camiones de carga, etapas a futuro contemplan la implementación del sistema IoT diseñado en diversas rutas y camiones, así como expandir los periodos de prueba para conformar un set de datos cuyo análisis futuro permita visualizar posibles implicaciones en la logística de comercio en transporte de carga terrestre.

## AGRADECIMIENTOS

## REFERENCIAS

Alla, H. R., Hall, R., & Apel, D. B. (2020). Performance evaluation of near real-time condition monitoring in haul trucks. *International journal of mining science and technology, 30*, 909–915.

Alquhali, A. H., Roslee, M., Alias, M. Y., & Mohamed, K. S. (2019). Iot based real-time vehicle tracking system. *2019 IEEE Conference on Sustainable Utilization and Development in Engineering and Technologies (CSUDET)*, (págs. 265–270).

Alrifaie, M. F., Harum, N., Othman, M. F., Roslan, I., & Shyaa, M. A. (2018). Vehicle detection and tracking system IoT based: a review. *Int. Res. J. Eng. Technol*, 1237–1241.

Arebey, M., Hannan, M. A., Basri, H., & Abdullah, H. (2009). Solid waste monitoring and management using RFID, GIS and GSM. *2009 IEEE Student Conference on Research and Development (SCOReD)*, (págs. 37–40).

Arguijo, J. E., León, E. G., Arellano, C. C., & García, F. R. (2019). Propuesta de sistema de gestión para optimización de redes de transporte público. *Res. Comput. Sci., 148*, 235–245.

Bahga, A., & Madisetti, V. (2014). *Internet of Things: A hands-on approach*. Vpt.

Balakrishna, S., & Thirumaran, M. (2019). Programming paradigms for IoT applications: an exploratory study. *Handbook of IoT and Big Data*, 23–57.

Beetrack. (July de 2019). 6 problemas de Distribución Logística de Productos [última Milla]. *6 problemas de Distribución Logística de Productos [última Milla]*. Obtenido de https://www.beetrack.com/es/blog/logistica-de-distribucion/

Bento, A. C. (2018). IoT: NodeMCU 12e X Arduino Uno, Results of an experimental and comparative survey. *International Journal, 6*.

Cadena, A. C., Matamoros, O. M., Pérez, D. A., & Escobar, J. J. (2019). Software de estación terrena para cohetes hidropropulsados. *Res. Comput. Sci., 148*, 305–322.

Calixto-Rodriguez, M., Valdez Martínez, J. S., Meneses-Arcos, M. A., Ortega-Cruz, J., Sarmiento-Bustos, E., Reyes-Mayer, A., . . . Domínguez García, R. O. (2021). Design and Development of Software for the SILAR Control Process Using a Low-Cost Embedded System. *Processes, 9*, 967.

Carbajal, J. (November de 2021). VMT inspecciona peso de los Camiones de Carga y a la vez realiza exámenes antidoping a motoristas. *VMT inspecciona peso de los Camiones de Carga y a la vez realiza exámenes antidoping a motoristas*. Noticias de El Salvador - La Prensa Gráfica | Informate con la verdad. Obtenido de https://www.laprensagrafica.com/elsalvador/VMT-inspecciona-peso-de-los-camiones-de-carga-y-a-la-vez-realiza-examenes-antidoping-a-motoristas-20211124-0062.html

Chanchí G, G. E., Ospina A, M. A., Campo M, W. Y., & others. (2021). IoT Architecture for Monitoring Variables of Interest in Indoor Plants. *Computación y Sistemas, 25*.

Cortes, R. (February de 2019). Soluciones Tecnológicas de Pesaje Inalámbrico en Vehículos de Carga. *Soluciones Tecnológicas de Pesaje Inalámbrico en Vehículos de Carga*. Obtenido de https://blogs.iadb.org/transporte/es/soluciones-tecnologicas-de-pesaje-inalambrico-en-vehiculos-de-carga/

de Corporin SA, C. V. (2021). Obtenido de http://www.corporin.com

DFRobot. (s.f.). Sim808withleonardomainboard. *Sim808withleonardomainboard*. Obtenido de https://wiki.dfrobot.com/SIM808-with-Leonardo-mainboard-SKU-DFR0355

Espino, L. E., Rios, Y. P., & Franco, E. G. (2020). Propuesta de un sistema de control, monitoreo y asistencia para optimización de recursos energéticos en el hogar. *Res. Comput. Sci., 149*, 49–61.

Espinoza, J. (March de 2018). Inseguridad en el norte de Centroamérica Afecta Al Transporte de Carga. *Inseguridad en el norte de Centroamérica Afecta Al Transporte de Carga*. El Nuevo Diario. Obtenido de https://www.elnuevodiario.com.ni/economia/420764-inseguridad-norte-centroamerica-afecta-transporte/

Feng, M. Q., & Leung, R. Y. (2020). Application of computer vision for estimation of moving vehicle weight. *IEEE Sensors Journal, 21*, 11588–11597.

Gallego Tercero, L. R., Menchaca Mendez, R., & Rivero Angeles, M. E. (2020). Spatio-Temporal Routing in Episodically Connected Vehicular Networks. *Computación y Sistemas, 24*.

Gohin Tay, C. A., & Vera Bernuí, K. E. (2015). Mejora del sistema de monitoreo y rastreo vehicular position logic-fermon Perú SAC. *Mejora del sistema de monitoreo y rastreo vehicular position logic-fermon Perú SAC*. Universidad Privada Antenor Orrego-UPAO.

Golondrino, G. E., Alarcón, M. A., & Muñoz, W. Y. (2020). Sistema IoT para el seguimiento y análisis de la intensidad de luz en plantas de interiores. *Res. Comput. Sci., 149*, 317–327.

Gómez, A. P., Cahuich, A. C., & Gómez, J. J. (2020). Plataforma de gestión IoT mediante técnicas de industria 4.0 para agricultura de precisión. *Res. Comput. Sci., 149*, 303–315.

Hejazi, H., Rajab, H., Cinkler, T., & Lengyel, L. (2018). Survey of platforms for massive IoT. *2018 IEEE International Conference on Future IoT Technologies (Future IoT)*, (págs. 1–8).

Hernandez, S., & Hyun, K. (2020). Fusion of weigh-in-motion and global positioning system data to estimate truck weight distributions at traffic count sites. *Journal of Intelligent Transportation Systems, 24*, 201–215.

Huang, C.-C., Lin, C.-L., Kao, J.-J., Chang, J.-J., & Sheu, G.-J. (2018). Vehicle parking

guidance for wireless charge using GMR sensors. *IEEE Transactions on Vehicular Technology, 67*, 6882–6894.

Jurado Murillo, F., Quintero Yoshioka, J. S., Varela López, A. D., Salazar-Cabrera, R., Pachón de la Cruz, Á., & Madrid Molina, J. M. (2020). Experimental Evaluation of LoRa in Transit Vehicle Tracking Service Based on Intelligent Transportation Systems and IoT. *Electronics, 9*, 1950.

Lv, W., Meng, F., Zhang, C., Lv, Y., Cao, N., & Jiang, J. (2017). A general architecture of IoT system. *2017 IEEE International Conference on Computational Science and Engineering (CSE) and IEEE International Conference on Embedded and Ubiquitous Computing (EUC), 1*, págs. 659–664.

Maia, J., & Yudi, J. (2020). An IoT solution for load monitoring and tracking of garbage-truck fleets. *2020 IEEE Conference on Industrial Cyberphysical Systems (ICPS), 1*, págs. 406–410.

Martínez, A., Onofre, H., Estrada, H., Torres, D., & Maquinay, O. (2018). Diseño y desarrollo de una arquitectura IoT en contexto con la plataforma FIWARE. *Res. Comput. Sci., 147*, 95–106.

Medel Juárez, J. d., Urbieta Parrazales, R., & Garduño Mendieta, V. (2019). Meteorological Portable System Consulted via Wi-Fi. *Computación y Sistemas, 23*.

Minero, E. (December de 2015). El Uso de Tecnología para aumentar La Productividad de la Carga. *El Uso de Tecnología para aumentar La Productividad de la Carga*. Obtenido de https://www.equipo-minero.com/contenidos/el-uso-de-tecnologia-para-aumentar-la-productividad-de-la-carga/

Moldtrans. (September de 2019). El Transporte de Mercancías Ha Mejorado, Pero Debe perfeccionarse. *El Transporte de Mercancías Ha Mejorado, Pero Debe perfeccionarse*. Obtenido de https://www.moldtrans.com/cuales-son-los-problemas-mas-habituales-en-el-transporte-de-mercancias/

Molina, Y. A., Ramírez, S. S., Morales, J. G., Reyes, A. M., Sánchez, R. G., & García, I. V. (2020). Diseño y desarrollo de un sistema de monitoreo remoto implementando Internet de las cosas. *Research in Computing Science, 149*, 235–247.

Moral, L. A. (2014). *Logistica del transporte y distribucion de carga*. Ecoe Ediciones.

Nayyar, A., & Puri, V. (2016). A review of Arduino board's, Lilypad's & Arduino shields. *2016 3rd international conference on computing for sustainable global development (INDIACom)*, (págs. 1485–1492).

Oskoui, E. A., Taylor, T., & Ansari, F. (2020). Method and sensor for monitoring weight of trucks in motion based on bridge girder end rotations. *Structure and Infrastructure Engineering, 16*, 481–494.

Polianytsia, A., Starkova, O., & Herasymenko, K. (2016). Survey of hardware IoT platforms. *2016 Third International Scientific-Practical Conference Problems of Infocommunications Science and Technology (PIC S&T)*, (págs. 152–153).

Putra, S. A., Trilaksono, B. R., Riyansyah, M., Laila, D. S., Harsoyo, A., & Kistijantoro, A. I. (2019). Intelligent sensing in multiagent-based wireless sensor network for bridge condition monitoring system. *IEEE Internet of Things Journal, 6*, 5397–5410.

Q. (October de 2020). El control satelital de camiones y por qué debes considerarlo. *El control satelital de camiones y por*

*qué debes considerarlo*. Obtenido de https://www.ubicalo.com.mx/blog/control-satelital-de-camiones/

Quijano, R. (November de 2018). Transporte de Carga: Un Trabajo de Peso: Noticias de El Salvador. *Transporte de Carga: Un Trabajo de Peso: Noticias de El Salvador*. El Diaro de Hoy. Obtenido de https://historico.elsalvador.com/historico/540323/transporte-de-carga-un-trabajo-de-peso.html

Radhakrishnan, K., Julien, C., Baranowski, T., O'Hair, M., Lee, G., De Main, A. S., . . . others. (2021). Feasibility of a Sensor-Controlled Digital Game for Heart Failure Self-management: Randomized Controlled Trial. *JMIR serious games, 9*, e29044.

Ray, P. P. (2016). A survey of IoT cloud platforms. *Future Computing and Informatics Journal, 1*, 35–46.

Rosa, V., & Flores-Cortez, O. O. (2017). Monitoreo remoto usando internet de las cosas. En IEEE (Ed.), *IEEE 37th Central America and Panama Convention (CONCAPAN XXXVII)* (págs. 1-3). Managua: IEEE. doi:10.1109/CONCAPAN.2017.8278466

Saleem, S. I., Zeebaree, S., Zeebaree, D. Q., & Abdulazeez, A. M. (2020). Building smart cities applications based on IoT technologies: A review. *Technology Reports of Kansai University, 62*, 1083–1092.

San Hlaing, N. N., Naing, M., & San Naing, S. (2019). GPS and GSM based vehicle tracking system. *International Journal of Trend in Scientific Research and Development (IJTSRD)*.

Sánchez, R., & Cipoletta Tomassian, G. (2003). *Identificación de obstáculos al transporte terrestre internacional de cargas en el MERCOSUR*. CEPAL.

Saritha, B., Bharadwaja, C. H., Nikhitha, M., Nethra Reddy, C. H., Arun, K., & Ahmed, S. M. (2022). An Intelligent Anti-Theft Vehicle Locking System Using IoT. En *ICDSMLA 2020* (págs. 1589–1595). Springer.

Seo, M. K., Shin, H. Y., Lee, H. Y., Ko, J. I., & Tumenjargal, E. (2021). Development of Onboard Scales to Measure the Weight of Trucks. *Journal of Drive and Control, 18*, 9–16.

Singh, K. J., & Kapoor, D. S. (2017). Create your own Internet of things: A survey of IoT platforms. *IEEE Consumer Electronics Magazine, 6*, 57–68.

Stawska, S., Chmielewski, J., Bacharz, M., Bacharz, K., & Nowak, A. (2021). Comparative accuracy analysis of truck weight measurement techniques. *Applied Sciences, 11*, 745.

Torres-Restrepo, L., Martínez-Rebollar, A., González-Mendoza, M., Estrada-Esquivel, H., & Vargas-Agudelo, F. (s.f.). Method for Introducing IoT Project Development Using Free Software Tools.

Villarreal, H., Lizarraga, M., Diaz-Ramírez, A., Rosas, V. Q., & García-Vázquez, J.-P. (2019). Componente del Internet de las cosas para detectar patrones de deambulaje en pacientes con demencia. *Res. Comput. Sci., 148*, 121–133.

Wipfli, B., Hanson, G., Anger, K., Elliot, D. L., Bodner, T., Stevens, V., & Olson, R. (2019). Process evaluation of a mobile weight loss intervention for truck drivers. *Safety and Health at Work, 10*, 95–102.

Zhao, Y., & Pan, Y. (2012). Electronic truck scale wireless remote control cheating monitoring system using the voltage signal. En *Advances in Electronic Commerce, Web Application and Communication* (págs. 323–326). Springer.